\documentclass{aa}

\usepackage{graphicx}
\usepackage{amsmath}	
\usepackage{amssymb}	
\usepackage{txfonts}
\usepackage{natbib}
\usepackage{longtable}
\usepackage[normalem]{ulem}
\usepackage{tablefootnote}

\usepackage{hyperref}
\hypersetup{
    colorlinks = true,
    citecolor ={blue}
}

\newcommand{\gaia}{$Gaia$\,}

\begin{document}

   \title{Volume-limited sample of low-mass red giant stars, the progenitors of hot subdwarf stars}

   \subtitle{l. Sample selection and binary classification method\thanks{Based on observations collected with the CORALIE echelle spectrograph on the 1.2-m Euler Swiss telescope at La Silla Observatory,
European Southern Observatory (ESO) through the Chilean Telescope Time under program ID's CN2019-58, CN2020B-77, CN2022A-82.}}

   \author{ Murat Uzundag\inst{1,2}, Mat\'ias I. Jones\inst{2}, Maja Vu\v{c}kovi\'{c}\inst{1}, Joris Vos\inst{3}, Alexey Bobrick\inst{4},  Claudia Paladini\inst{2}
   }
           \institute{Instituto de F\'isica y Astronom\'ia, Universidad de Valpara\'iso, Gran Breta\~na 1111, Playa Ancha, Valpara\'iso 2360102, Chile
           \\
           \email{murat.uzundag@postgrado.uv.cl}
           \and
           European Southern Observatory, Alonso de Cordova 3107, Santiago, Chile
           \and
            Astronomical Institute of the Czech Academy of Sciences, CZ-25165, Ond\v{r}ejov, Czech Republic
            \and
            Technion - Israel Institute of Technology, Physics Department, Haifa, Israel 32000
\\
        }
\date{}

    \titlerunning{Volume-limited sample of low mass red giant stars}
	\authorrunning{Murat Uzundag et al.}

  \abstract
   {The current theory predicts that hot subdwarf binaries are produced from evolved low-mass binaries that have undergone mass transfer and drastic mass loss during either a common envelope phase or a stable Roche lobe overflow while on the red giant branch (RGB).
   }
   {
   We perform a spectroscopic survey to find binary systems that include low-mass red giants near the tip of the RGB, which are predicted to be the direct progenitors of subdwarf B (sdB) stars. We aim to obtain a homogeneous sample to search for the observational evidence of correlations between the key parameters governing the formation of sdB stars and constrain the physics of stable mass transfer.
   }
   {Based on data from the $Gaia$ mission and several ground-based, multi-band photometry surveys, we compiled a sample of low-mass red giant candidates.
   The candidates were selected according to their $Gaia$ data release 2 (DR2) color, absolute magnitude and proper motion cuts.
   In this work, we concentrated on the southern hemisphere targets and conducted a spectroscopic survey of 88 red giant stars to search for the long-period RGB 
   + MS binary systems within 200\,pc. 
   Combining radial velocity (RV) measurements from ground-based observations with CORALIE and RV measurements from $Gaia$ DR2 and early data release 3 (eDR3) as well as the astrometric excess noise and RUWE measurements from $Gaia$ DR3, we defined a robust binary classification method. In addition, we searched for known binary systems in the literature and in the $Gaia$ DR3.}
   {
   We select a total of 211 RGB candidates in the southern hemisphere within 200\,pc based on the \gaia DR2 color-magnitude diagram. Among them, a total of 33 red giants were reported
   as binary systems with orbital periods between 100 and 900 days, some of which are expected to be the direct progenitors of wide binary sdB stars. 
   In addition, we classified 37 new  MS\,+\,RGB binary candidates, whose orbital parameters will be measured with future spectroscopic follow-up.
   }
   {Using high-quality astrometric measurements provided by the $Gaia$ mission coupled with high-resolution spectroscopy from the ground, we provide a powerful method to search for low-mass red giant stars in long-period binary systems. }


   \maketitle

\section{Introduction}


Hot subdwarf B stars (sdBs) are evolved low-mass red giants that have had almost all of their envelope stripped away near the tip of the first red giant branch (RGB), according to standard evolutionary scenarios \citep{dorman1993}.
They are evolved, compact ($\log{g}$ = 4.5 - 6.5\,dex) and hot ($T_{\rm eff}$ = 20\,000 - 40\,000\,K) objects with radii between 0.15 $R_{\odot}$ and 0.35 $R_{\odot}$, located on the so-called extreme horizontal branch \citep[EHB; see][for a review]{heber2016}.
Their characteristic mass is close to the core-He-flash mass ($\sim$0.47\,$M_{\odot}$) and they have a very thin hydrogen envelope (M$_{\rm{H}}$\,$<$\,0.01\,$M_{\odot}$). 
It was found that a large proportion of sdBs (40–70\%) are in binaries \citep[e.g.][]{maxted2001,Napiwotzki2004,Cop2011}, implying that binary interactions are the most likely explanation for their formation.



The three binary formation channels that contribute to the sdB population are: the common-envelope (CE) ejection channel \citep{Paczynski1976,han2002}, the stable Roche-lobe overflow (RLOF) channel \citep{han2002,han2003} and a binary white-dwarf (WD) merger \citep{webbink1984}. 
The sdB component in both binary scenarios (CE and RLOF) has a relatively narrow mass range, peaking at 0.47 $M_{\odot}$, whereas the double helium WD merger channel creates a single subdwarf star with a relatively wide and flat distribution, from 0.42 up to 0.72 $M_{\odot}$ \citep{han2003}.
\citet{han2002} showed that the CE ejection channel leads to close binaries with WD or main sequence (MS) companions with short orbital periods ranging from hours to tens of days.
Much of the observational research has focused on these short-period systems, and currently, more than 300 short-period sdB systems have been detected \citep[references therein]{Kupfer2015, Sc2019, Dai2022}. 

The RLOF channel generates sdB+MS binaries with orbital periods ranging from  400 up to 1600 days based on the the binary population synthesis models \citep[see red dashed line in Fig. 3 of][]{VosJ2019}.
These systems are challenging to detect, given their long orbital periods.
Long-term observational campaigns have been dedicated to discover these systems over the last decades.  
Recently, the orbital parameters of long-period sdB binaries have been found for 26 systems \citep{VosJ2019,otani2021,nemeth2021, molina2021} with periods ranging from  400 to 1600 days. 
Because both stars are visible in the spectrum, these systems are referred to as "composite" binaries which account for 30–40\% of all sdBs \citep{VosJ2019}.
The observed composite binary systems present correlations between orbital elements, which revealed an invaluable source of information to test the RLOF channel.
One of the first correlations that have been found is that between eccentricity and period in which the eccentricity increases with increasing orbital period \citep{VosJ2015}. 
This finding contradicts evolutionary models since all current models predict circular orbits.
\citet{VosJ2015} demonstrated that the observed eccentricities of wide sdB binaries may be explained by combining two eccentricity pumping mechanisms, phase-dependent RLOF and a circumbinary disk. These models, on the other hand, fail to replicate the observed trend of greater eccentricity with longer orbital periods.
Furthermore, the wide sdB binaries show another correlation between the mass ratio (q) and the orbital period ($P_\mathrm{orb}$) \citep{VosJ2019}.
The chemical history of our Galaxy can precisely explain the observed relationship between the mass ratio and the orbital period in long-period sdB binaries as found by \citet{VosJ2020}.
The authors performed a small but statistically significant binary population synthesis study with the binary stellar evolution code MESA considering 
binaries with primary masses between 0.7\,$M_{\odot}$  and 2.0\,$M_{\odot}$ and initial periods between $\sim$100 and $\sim$900 days.
The authors find three important correlations between the properties of the sdB + MS binaries and their progenitors.  The RG mass is inversely correlated with the mass ratio of the sdB + MS system, the RG + MS orbital period is correlated with the sdB + MS mass ratio, and lastly the metallicity is correlated with the MS mass in the sdB + MS systems. They also provide ranges on the orbital periods and mass ratios of RG + MS systems that can form wide sdB binaries. By studying the properties of the RG + MS binaries, we can verify if the population of RG + MS binaries matches the requirements needed to produce the observed number of wide sdB binaries. Furthermore, since several properties are linked directly between the progenitors and the final sdB binaries, knowing the distribution of those parameters in the RG + MS binaries will directly predict the expected observed population, and will allow for a strict test of the proposed formation channel.

The aim of this study is to analyze the observed characteristics of a statistically significant volume-limited sample of low mass evolved RGB binaries that are potential progenitors of the wide sdB binaries. 
We perform a spectroscopic survey searching for binarity among selected low-mass RGB candidates. 
In future, obtained orbital solutions of RGB+MS binaries initiated by this work will be quite useful for comparing the population of progenitors with the population of the long-period sdB binaries. 
The observed volume-limited sample composed of pre- and post-mass transfer binaries will provide additional information about the details of mass transfer in these systems, as well as the nature of the period-eccentricity relation observed in long-period systems.


The paper is  organized as  follows.  In Section \ref{samle_selection}, 
we provide the sample selection from $Gaia$ DR2 and the literature. In  Sect. \ref{observations}, we
describe the EULER/CORALIE observations and data reduction.
In Sect. \ref{CM}, we describe our classification method to search for low-mass red giants in long-period binary systems.   
Finally, in Sect. \ref{conclusion}, we summarize our results and give an outlook for the future.

\section{Target selection} 
\label{samle_selection}

To select the low-mass red giant candidates, we have used the $Gaia$
mission data release 2 catalog \citep[DR2;][]{DR22018} in combination with synthetic colors for RG stars obtained from the MESA Isochrones and Stellar Tracks (MIST; \citealt{Choi2016}).
First, we applied the criteria on photometric and astrometric quality as described in \citet[][Appendix C]{Lindegren2018}. 
Then, we chose objects in the region defined by the four functions below:
\begin{equation}
G_{\rm abs} = 2.19 (G_{BP} - G_{RP}) - 1.95   ;
\end{equation}

\begin{equation}
G_{\rm abs} = 1.86 (G_{BP} - G_{RP}) - 1.10   ;
\end{equation}

\begin{equation}
G_{\rm abs} = 45 (G_{BP} - G_{RP}) - 43.9   ;
\end{equation} 

\begin{equation}
G_{\rm abs} = 28.33 (G_{BP} - G_{RP}) - 38.28   ;
\end{equation}
 where $G_{\rm abs}$ is an absolute $Gaia$ G magnitude. Since ${\tt phot_{-}g_{-}mean_{-}mag}$ of $Gaia$ is apparent magnitude, $G_{\rm abs} = {\tt phot_{-}g_{-}mean_{-}mag} + 5 \log_{10}({\tt parallax}/1000) + 5$.
These cuts focus on the systems prior to the tip of the RGB, and cut off evolved higher-mass systems.
 Equations 1 and 4 provided the cuts to exclude the contaminants from the subgiant and horizontal branch stars, while equations 2 and 3 provided the cuts to remove the contaminants from the red clump stars. 
The selected region is shown in Fig. \ref{fig:CMD_sd} (left panel) as a red shaded region, which includes a total of 6158 targets.
We did an external cross-match with several catalogues using TOPCAT \citep{taylor2005}. 
The literature was searched through the SIMBAD Astronomical Database\footnote{https://simbad.u-strasbg.fr/} \citep{Wenger2000} and we removed 158 outlier objects which turn out to be various types of systems from long period variables (LP) to neutron stars ($N^{*}$).
The remaining 6000 targets were then cross-matched with the other large-area surveys using a 5 arcsec aperture in order to clean the sample further. 
We cross-matched the data with the GALEX DR5 All-sky Imaging Survey \citep[AIS;][]{Bianchi2011} and found 3331 objects with NUV and FUV measurements.
In order to obtain the near-IR measurements, we cross-matched these objects with the Wide-field Infrared Survey Explorer \citep[WISE;][]{wright2010}, and found 3325 targets with their W1 and W2 band measurements. 
Likewise, the other IR surveys, including Infrared Astronomical Satellite \citep[IRAS;][]{N1984} and the Infrared Astronomical Mission \citep[AKARI;][]{Murakami2007} were cross-matched to exclude the targets further if they show any excess. 
The UV excess indicates WD companions while the IR-excess indicates the presence of disks or dust.  
These kinds of systems (548 stars) were excluded from the survey since we are searching for RG + MS stars that have not yet undergone any type of binary interaction.
Finally, we selected 2777 low-mass RG candidates within 500\,pc covering the southern and northern hemispheres. All selected objects have an error on the parallax lower than 10 \%.

In this study, we limited ourselves only to the southern hemisphere ($\delta \leq 20^{\circ}$) and to the smaller volume of 200\,pc so that we have a feasible number of systems for a pilot study. Within 200\,pc, we have a total of 211 low-mass RG candidates in the southern hemisphere.
In Fig. \ref{fig:CMD_sd}, we present the color-magnitude diagram of selected low-mass RG candidates within 200\,pc with blue dots (left panel) and their spatial distribution 
with respect to the galactic coordinate system (right panel).

\subsection{Spectroscopic binaries from the literature}
 
We searched for known low-mass RGs from the literature applying the similar selection criteria as described in Sect. \ref{samle_selection}. 
We found a total of 300 binaries detected by different spectroscopic surveys \citep[e.g.][]{Setiawan2004,Jones2011, Massarotti2008,Wittenmyer2011}. 
Among them, we identified 24 RG binary systems with orbital periods ranging from 100 to 900 days, which could be potential progenitors of wide sdB + MS composite systems produced by RLOF mechanism.  
The orbital elements of these systems are listed in Table \ref{tab:known_systems} including the orbital period of the binary system (P), the amplitude of the radial velocity curve (K),  the eccentricity of the orbit ($e$), angle of periastron ($\omega$), time of periastron (T$_{\rm 0}$) and the mass function (f(m)).

\subsection{Binaries from DR3}
We made use of Non-single stars catalog (I/357\footnote{\url{https://cdsarc.cds.unistra.fr/viz-bin/cat/I/357}}) from $Gaia$ DR3 \citep{DR32022} to detect more binary stars within 200\,pc. 
To select all RG candidates in binary systems, we used the same selection criteria described in Sect. \ref{samle_selection}.
DR3 catalog includes astrometric, spectroscopic, and eclipsing binaries (where solutions from the combinations of astrometry and radial velocities, or eclipsing binary light curves and radial velocities are also provided; \citealt{DR32022}).
From the $Gaia$ DR3 catalog, we found nine binary systems with orbital periods between 100 and 900 days. All these systems and their orbital elements are presented in Table \ref{tab:DR3_binaries}.

\begin{figure*}
    \centering
    \includegraphics[clip=,width=\textwidth]{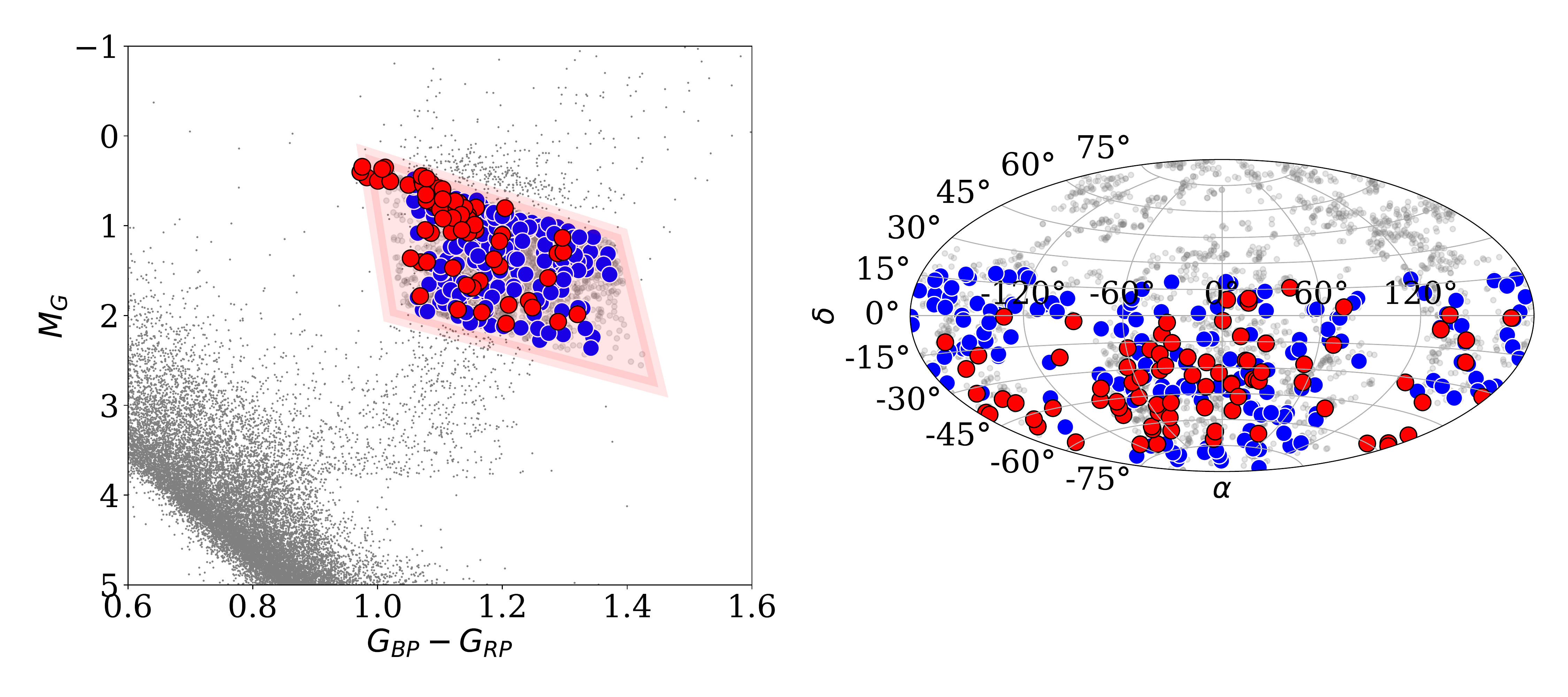}
     \caption{{\sc Left:} The color-magnitude diagram of all low-mass  ($0.7 - 2.3 M_{\odot}$) RG candidates within 200 pc (blue dots) from $Gaia$ DR2, including only the southern hemisphere stars.
     As a comparison, the $Gaia$ DR2 color-magnitude diagram of the 100 pc clean sample \citep[Sample C in][]{Lindegren2018} is shown in grey.
      The slightly larger grey dots within the selected sample represent the low-mass RG candidates within 500 pc.
     The red circles show the low-mass RG candidates that were observed with EULER/CORALIE.
     {\sc Right:} Sky locations (Galactic coordinates, Aitoff projection) of the volume-limited low-mass RG sample within 200\,pc with respect to the galactic coordinate system using the same color coding. }
    \label{fig:CMD_sd}
\end{figure*}

\begin{table*}
\setlength{\tabcolsep}{10pt}

\caption{Orbital solutions for all known binaries, including the Hipparcos catalog number, the orbital period of the binary system (P), the amplitude of the radial velocity curve (K),  the eccentricity of the orbit ($e$), angle of periastron $\omega$, time of periastron (T$_{\rm 0}$) and the mass function (f(m)).
}
\begin{tabular}{cccccccccc}
\hline \hline
HD	& HIP &   P 	&  K  &  $e$	  &  $\omega$	  &  T$_{\rm 0}$  &  f(m)  & Ref. \\ 
    &     &     (day)        &     (km\,s$^{-1}$)              &           &               &               & (M$_{\odot}$)  &\\
\hline
5516    & 4463   &  115.7 (0.2)  &   17.91   (0.09)   & 0.003      & 103              &  52662             &  3.3127          &  2   \\
5877    & 4618   &  211.4   (0.3)     &   12.94 (0.01) & 0.1   & 40.8 (0.6) &  5406.3 (0.3)&  0.467           &  3   \\
14355   & 10548  &  429.1   (0.3)     &   6.9  (0.3)& 0.3   & 5.0 (1.9)  & 5306.6 (1.5) & 12.9 (0.2)  &  3   \\
15755	& 11840  &  629.2   (2.7)      &   10.67               & 0.14       & 294.6            &  53979             &  0.077           &  2   \\
27697   & 20455  &  522.1   (1.8)      &   2.84    (0.3)  & 0.48        &                  &                    &  0.0084          &  1   \\
29923	& 22055  &  680.1  (25.9)    &   6.48                & 0.1       & 42               &  53797             &  0.0189          &  2   \\
30197	& 22176  &  107.6  (0.1)    &   8.51    (0.15)  & 0.03      & 254.5            &  49457.5           &  0.00623         &  2   \\
32008	& 23221  &  898.1   (2.7)      &   5.2                 & 0.17       & 155.1            &  51240             &  0.01248         &  2   \\
54563	& 34608  &  113.4	(0.1)  &   20.75   (0.04)   & 0.4	        &                  &                    &                  &  2   \\
62644   & 37606  &  380.6   (0.1)      &   9.84    (0.7)  & 0.73        &                  &                    &  0.12            &  1   \\
65938	& 39198  &  365.4  (0.6)    &   9.56                & 0.52       & 108.1            &  53826.9           &  0.0208          &  2   \\
72184	& 41935  &  324	1  (16.5)   &   3.83                & 0.15       & 90               &  54051             &  0.00183         &  2   \\
94386   & 53259  &  925     (1)  &           2.18   (0.01)   & 0.42    & 37.8 (0.2)     & 54582.1 (1.2) &               &  2,4   \\
102928  & 57791	 &  489.6  (0.9)    &   12.48   (0.2)  & 0.22        & 102.4            &  53114.4           &  0.0916          &  2   \\
104358  & 58601  &  281.1   (0.3)  &       1.83   (0.04)   & 0.24   & 146 (7)     & 54836.8 (4.2) &                  &  4   \\
133166  & 73758 &    97.1   (0.2)  &     11.82 (0.01)   & 0.4  & 53.4 (0.1) &  5304.3 (0.1)&  0.271           &  3   \\
136138	& 74896  &  508.7   (1.6)      &   6.19    (0.12)  & 0.33       & 39               &  53751.8           &  0.01053         &  2   \\
153438  & 83224 &   173.3   (0.3)  &      8.74  (0.01) & 0.3   & 85.6 (0.3) &  5251.3 (0.1) &  0.104           &  3   \\
153956  & 83138	 &  900.14  (14)      &   3.36    (0.16)  & 0.07        & 338              &  52956             &  0.00352         &  2   \\
172831	& 91751  &  485.3   (0.3)      &   9.68                & 0.21       &                  &                    &                  &  2   \\
179799  & 94521  &  856.1   (39.1)     &   5.99    (0.4)  & 0.66        &                  &                    &  0.081           &  1,2 \\
181391	& 95066  &  266.5 (0.1)  &   29.86               & 0.83       &                  &                    &                  &  2   \\
199870	& 103519 &  635.1   (0.5)      &   6.44                & 0.44       &                  &                    &                  &  2   \\
221625	& 116243 &  94.9  (0.1)   &   36.21               & 0.52     & 315.82           &  53555.4           &  1.67            &  2   \\

\hline
\end{tabular}
\tablebib{
(1)~\citet{Setiawan2004}; (2) \citet{Massarotti2008}; (3) \citet{Bluhm2016}; (4) \citet{Wittenmyer2016}
}
\label{tab:known_systems}
\end{table*}

\begin{table}
\renewcommand{\arraystretch}{1.1}
\setlength{\tabcolsep}{2.8pt}

\centering
\caption{Binary systems that are found in DR3, including the name of star, the methods that have been used (astrometric $+$ spectroscopic,	astrometric and	astrometric, respectively), the orbital period of the system, time of periastron (T$_{\rm 0}$) and eccentricity of the orbit. 
} 
\centering
\begin{tabular}{lcccc}
\hline\hline
\vspace{-0.3cm} \\
Star         &   Solution type &  Period & T$_{\rm 0}$  & e       \\
             &            & (days)  &  (days)      &            \\
\hline \vspace{-0.3cm} \\

HD2132   &	AstroSpectroSB1     	&  773	  &    369.636	 &   0.04       \\
HD8410   &	AstroSpectroSB1     	&  304	  &    30.240	 &   0.22       \\
HD9525   &	Orbital             	&  582	  &    46.125	 &   0.16       \\
HD36787  &	AstroSpectroSB1     	&  257	  &    108.608	 &   0.42       \\
HD116338 &	OrbitalTargetedSearch	&  248	  &    -34.041	 &   0.84       \\
HD13423  &	OrbitalTargetedSearch	&  899	  &    303.991	 &   0.21       \\
HD162049 &	OrbitalTargetedSearch	&  713	  &    -83.897	 &   0.45       \\
HD190574 &	AstroSpectroSB1     	&  552	  &    -90.773	 &   0.70       \\
HD202470 &	AstroSpectroSB1     	&  488	  &    241.819	 &   0.52       \\


\hline 
\label{tab:DR3_binaries}
\end{tabular}
\end{table}


\begin{figure*}
    \includegraphics[clip,width=1.0\columnwidth]{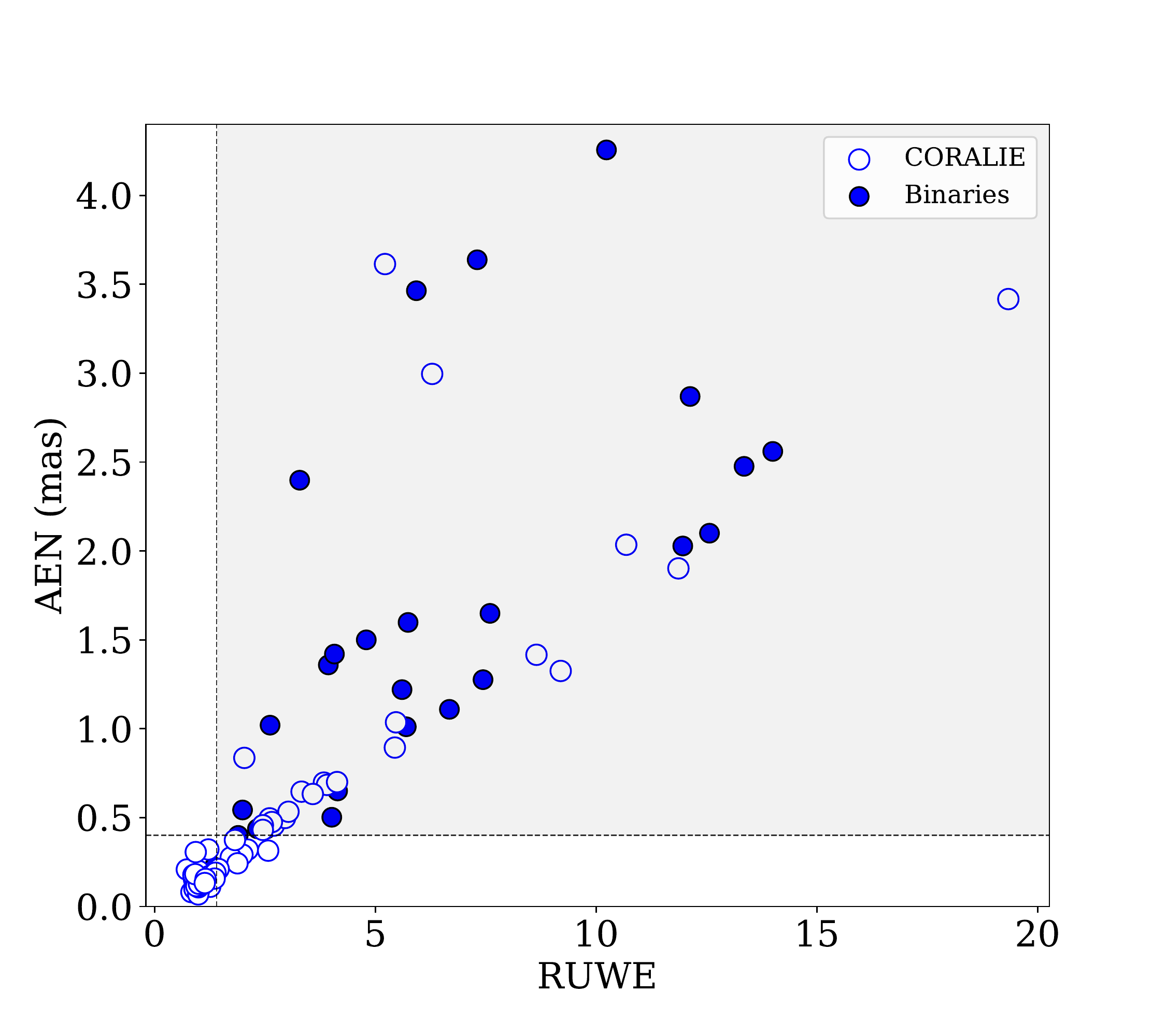} 
    \includegraphics[clip,width=1.0\columnwidth]{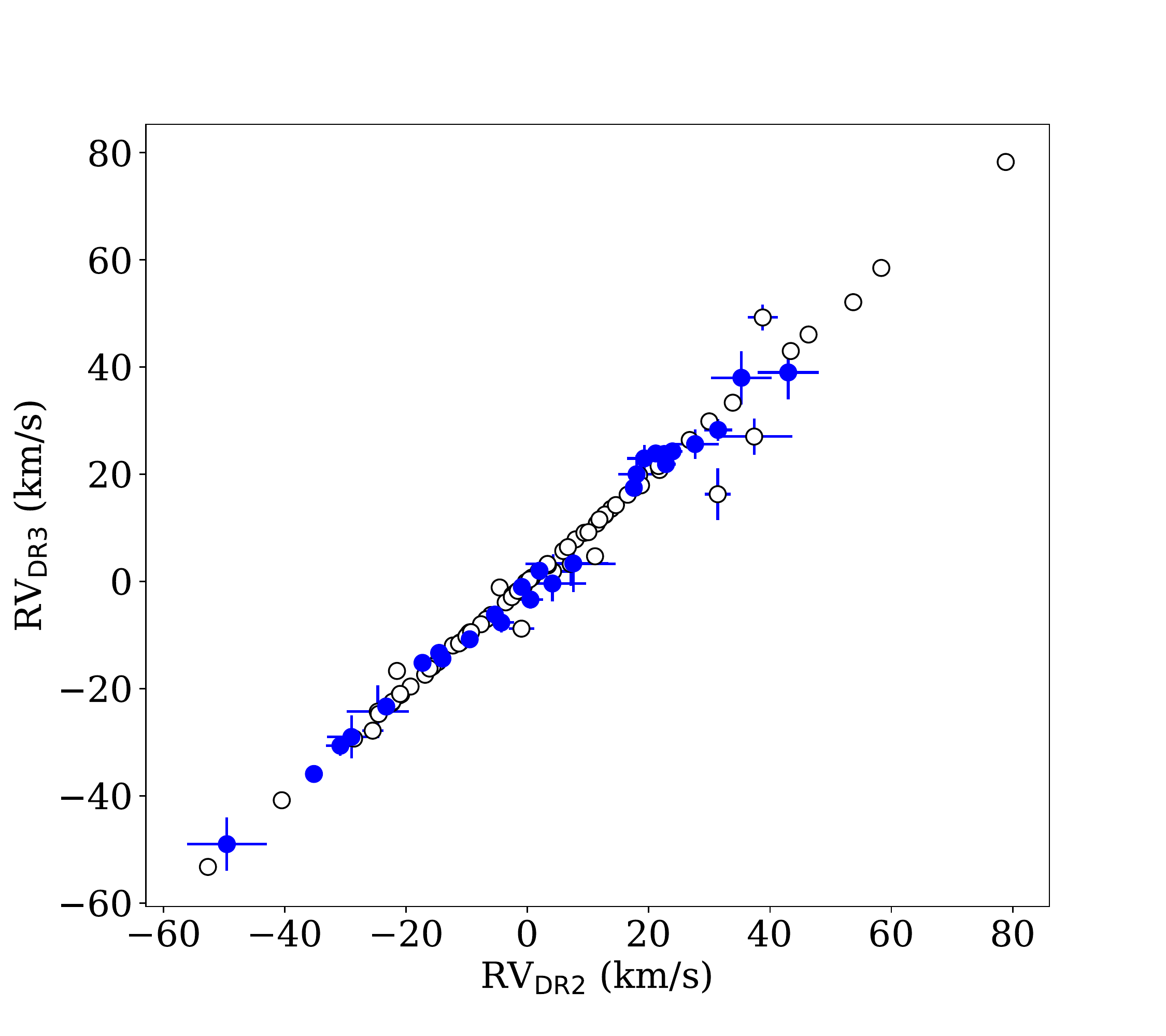} 
 \caption{{\sc Left:} AEN versus RUWE, including known binaries from the literature and DR3 (filled blue circles) and observed stars (open blue circles). 
 AEN of 0.4 mas is represented by the horizontal black dashed line, and RUWE of 1.4 is represented by the vertical black dashed line. The gray-shaded area displays all the stars that are most likely in binary systems.
         {\sc Right:} Radial velocity measurements from DR3 (y-axis) as a function of radial velocity measurements from DR2 (x-axis) using the same color coding.}
    \label{fig:AEN_ruwe_rv}
\end{figure*}

\section{Observations and data reduction}
\label{observations}

\subsection{Observations and data reduction using CORALIE}

The spectroscopic observations of the low-mass RGB candidates analyzed in this paper were 
obtained with the CORALIE echelle spectrograph \citep{Queloz2001} mounted at the Swiss 1.2-metre Leonhard Euler Telescope at La Silla Observatory in Chile.
CORALIE has a resolving power of R\,$\sim$60000, allowing for a long-term radial velocity precision up to 0.03 km\,s$^{-1}$ \citep{Queloz2001}.
CORALIE is fed by two fibers: a $2''$ diameter on-sky science fiber encompassing the star, and another that can either be connected to a Fabry-P\'erot etalon for simultaneous wavelength calibration (used in the case of our survey) or on-sky for background subtraction of the sky flux.


From the 200\,pc RG sample, we observed 82 stars with Euler/CORALIE. 
The observations were done in three different runs.
The details of these observations are given in Table \ref{tablespec1} including the instrument, date, spectral range and the signal-to-noise (S/N) ratio. 
The observed stars have a photometric $G$ mean magnitude ranging from 5 to 9 with a median of 7.5.
The observed sample is depicted in Fig. \ref{fig:CMD_sd} (red dots).
In total, we obtained 123 high-resolution CORALIE spectra for 82 stars.
We reduced and analyzed the data using the customized $\tt CERES$ pipeline \citep{Brahm2017}, which performs all the extraction processes from basic bias, dark and flat-field corrections (including scattered light) to order tracing, wavelength calibration, and computation of precise radial velocities using the cross-correlation technique.
For 38 objects, we obtained two epochs spectra and measured their RV shifts.
The RV measurements from multi-epoch CORALIE observations are listed in the tables in the Appendix \ref{Appendix}. 
The difference between these two different epochs is defined as $\Delta$CRV and is provided in Table\,\ref{tab:C2019}.

\subsection{ESO archival data}

We used the ESO archive\footnote{\url{http://archive.eso.org/eso/eso_archive_main.html}} to search for high-resolution spectra of stars in our sample.
We found a total of 8 stars observed with HARPS and 19 targets observed with FEROS. 
However, most of them (18 stars) were observed by the EXPRESS RV program \citep{Jones2011, soto2021}, whose spectroscopic binaries  were presented in \citet{Bluhm2016} and are listed in Table \ref{tab:known_systems}. 
One star is in a WD+AFGK binary system \citep{Ren2020} and two objects are found to be single planet-host stars \citep{Yilmaz2017,Witten2020}, which are discarded from our analysis. 
Finally, our analysis includes six stars observed with FEROS.
The $\tt CERES$ pipeline was used to reduce the FEROS spectra as it was for CORALIE data.

\begin{table}
\setlength{\tabcolsep}{2.2pt}
\renewcommand{\arraystretch}{1.1}
\centering
\caption{Observing log of the spectroscopic data obtained for the low-mass red giant stars studied in this work. 
}
\begin{tabular}{cccc}
\hline \hline
  Instrument & Date   & Range & S/N \\
               &        &  ($\AA$) &   \\
\hline

  CORALIE      &  15-16-17 June 2019        &   3800-6800   &  30-80    \\
  CORALIE      &  18-19-20 November 2021    &   3800-6800   &  50-100   \\
 CORALIE      &  4 April 2022              &   3800-6800   &   40-60   \\

\hline 
\label{tablespec1}
\end{tabular}
\end{table}

\section{Classification Method}
\label{CM}

\subsection{Astrometric excess noise}

The high astrometric precision currently made possible by missions such as $Gaia$ provides several important parameters.
One of the most important parameters for this work is the \gaia DR2 astrometric excess noise (AEN), which is a measure of the residuals in the source's 5-parameter astrometric solution. It can be used in astrometric binary system classification since a higher AEN value might indicate the presence of unresolved companions. \citet{Gandhi2022} reviewed the details and feasibility of this method, as well as an effective way to use AEN to search for X-ray binaries.

For our study, we made use of AEN to search for low mass RG binary candidates in our sample. 
Starting from the 33 known RG binaries (from the literature and DR3), we analyzed their AEN measurements. 
Out of 33 known binaries 31 have AEN $\geq$ 0.4\,mas (see Fig. \ref{fig:AEN_ruwe_rv} filled circles). Given that more than 90\% of the known binaries have AEN $\geq$ 0.4\,mas, we have chosen this value of AEN as a threshold to consider that a star is a potential binary candidate. 
In other words, these targets will be further followed up in order to confirm the binarity and finally obtain orbital periods. 

\subsection{The Renormalised Unit Weight Error (RUWE)}

The $Gaia$ data product known as the Renormalised Unit Weight Error (RUWE) is a measure used to filter out sources with potentially bad astrometry, in particular spurious parallaxes or proper motions \citep[see for details,][]{Lindegren2018}.
The RUWE is expected to be around 1.0 for sources where the single-star model provides a good fit to the astrometric observations. 
If a value is significantly greater than 1.0, that could indicate that the source is not a single object or otherwise problematic for the astrometric solution.
Therefore, we used RUWE to search for low-mass RG binary candidates in our sample.
We examined the RUWE measurements of the known RG binaries, and we found that 93\% of the known binary systems have RUWE bigger than 1.4.
Therefore we used RUWE $\geq$ 1.4 as a threshold to consider the object as a binary candidate.

To put this in a perspective, we have plotted known binary systems as well as observed stars with CORALIE and FEROS in AEN and RUWE parameters space indicating our selection threshold in Fig. \ref{fig:AEN_ruwe_rv}. The gray shaded area depicts the systems that are potential binaries based on our selection criteria. As can be seen in Fig. \ref{fig:AEN_ruwe_rv}, all known binaries fall on the grey shaded area. 

\subsection{Radial velocities from CORALIE, FEROS, DR2 and DR3 \label{sec:RVs}}

In order to define the binary fraction in our observed sample, we made use of the radial velocity measurements from CORALIE, FEROS, $Gaia$ DR2 and DR3.
The RV measurements from DR2 and DR3 are depicted in Fig. \ref{fig:AEN_ruwe_rv} (right panel), where we show known binaries with filled blue circles and observed stars with open blue circles. As can be seen in Fig. \ref{fig:AEN_ruwe_rv} (right panel), DR2 and DR3 measurements do not differ. 

With CORALIE, we obtained two different epochs of spectra ($\Delta$CRV) for 38 stars, which are reported in Table \ref{tab:C2019} where we also listed the AEN value and the corresponding uncertainty. 
If the difference in $\Delta$CRV $\geq$ 0.1 km\,s$^{-1}$, we consider it as a potential binary candidate. 
Out of these 38 observed objects, we found 17 stars that show $\Delta$CRV above the threshold and, therefore, could be potential binary systems.  

We obtained a single epoch CORALIE spectrum for 44 stars, which were combined with RV measurements from $Gaia$ DR2 in order to measure RV differences. 
Likewise, a single epoch archival FEROS spectra for 6 stars were combined with RV measurements from $Gaia$ DR2.
In Table \ref{tab:C20191}, we show the difference between the CORALIE/FEROS RV and the DR2 RV measurements ($\Delta$RV) of all the single epoch observed stars from the ground together with their AEN and RUWE. 
Given that the errors in $Gaia$ DR2 measurements vary from 0.2 to 2.5 km\,s$^{-1}$, we have chosen to use the 3\,$\sigma$ criteria as the threshold. In other words, if the RV difference between ground base and DR2 RV measurements is three times greater than its RV error, we consider the object as a potential binary candidate. 

\subsection{Classification}

By combining three sets of measurements, AEN, RUWE and RV measurements, we defined three categories that classify a star by its likelihood to be a binary candidate.
The stars that have been observed from the ground are divided into three categories. 
The stars that have at least two of the parameters (AEN/RUWE and RV variation) larger that the threshold we classify as binary candidates and assign them a category 1. The stars that have one of the parameters above the threshold  (AEN,  RUWE or RV variation) we assign them category 2 as they are likely to be binary candidates.
The third group (category 3) includes targets with no significant AEN, RUWE or large RV variations.

In the observed set, we found that 24 stars show AEN $\geq$ 0.4, out of which 18 stars also present significant radial velocity variations. It is these 18 stars that we assigned a category 1 as they are binary candidates which will be priority targets for follow-up observations.
Furthermore, we found 18 stars that are potential binary candidates (category 2) out of which 13 
targets show only significant RV variations, and  5 stars show solely astrometric excess noise above the threshold of 0.4\,mas. These 18 stars are also added to the follow-up observations as additional observations are needed to confirm their binarity.  

Furthermore, three stars (HD119483, HD149649 and CD-397574) that were observed with CORALIE in two epochs show RV variations of 0.061, 0.046 and -0.046 km$s^{-1}$, respectively. As these variations are just below the RV threshold of 0.1 kms$^{-1}$ we marked these stars as  potential follow-up candidates. 
Moreover there is one target, HD120144, that has RUWE of 1.37, while it does not show either AEN or RV variation. We consider this target as a possible binary candidate (category 2), which should also be observed in the future. 


Lastly, using the same classification method, we classified 6 FEROS targets that were obtained from the ESO archive. One of them corresponds to a binary system listed in the \gaia DR3 catalogue, and one more was classified in category 1, i.e., it is most likely a binary candidate.


\begin{table}
\renewcommand{\arraystretch}{1.1}
\setlength{\tabcolsep}{2pt}
\centering
\caption{
Binary classification of low-mass red giant stars reported in this paper with two CORALIE epochs spectra, including the name of the observed star, the astrometric excess noise (AEN) measurements from $Gaia$ eDR3, the difference between two CORALIE RV measurements ($\Delta$CRV), the errors on RV and the classification number.
}
\centering
\begin{tabular}{lcccccc}
\hline\hline
\vspace{-0.3cm} \\
Star         &     $\Delta$CRV    & eRV            &  AEN    &  RUWE         & Category  \\
             &   (km\,s$^{-1}$)  &  (km\,s$^{-1}$)  &(mas)   &           &  \\
\hline \vspace{-0.3cm} \\
HD112521      &   3.715	     &  0.04	& 0.68     &   3.899      &  1       \\
HD167936  	&   -16.23       &  0.05	&	1.42   &   8.646      &  1       \\
HD305357  	&    4.276	     &  0.1	    & 0.46     &   2.447      &  1       \\
HD214941  	&  3.44          &  0.06    &	0.43   &   2.446      &  1       \\
HD217614  	&  -10.76        &  0.05    &	2.04   &   10.68      &  1       \\
HD201013  	&  -5.57         &  0.04    &	0.7    &   4.128      &  1       \\
HD206005  	&   -0.49        &  0.03    &	0.37   &   1.814      &  1       \\
HD71464   	&   22.710	 	 &  0.04	&	0.15   &   1.285      &  2       \\
HD116338  	&   1.358	     &  0.03    &	0.21   &   1.445      &  2       \\
HD136350  	&   -9.882	     &  0.05	&	0.09   &   0.998      &  2       \\
HD139137  	&   0.955	     &  0.06	&	0.31   &   2.567      &  2       \\
HD204381  	&   -0.012	     &  0.04	&	0.84   &   2.027      &  2       \\
HD85885   	&   -0.198	     &  0.04	&	0.24   &   1.872      &  2       \\
HD102805  	&   0.518	     &  0.04	&	0.09   &   0.924      &  2       \\
HD170105  	&   1.067	     &  0.04	&	0.1    &   0.897      &  2       \\
HD157527  	&   -0.306	     &  0.04	&	0.21   &   0.724      &  2       \\
HD99891   	&   -0.136	     &  0.04	&	0.17   &   1.352      &  2       \\
HD120144  	&   0.013	     &  0.04	&	0.19   &  1.373       &  2       \\
HD119483  	&   0.061	     &  0.04	&	0.15   &   1.207      &  3       \\
HD149649  	&   0.046	     &  0.04	&	0.11   &   0.954      &  3       \\
CD-397574 	&   -0.046	     &  0.05	&	0.08   &  0.984       &  3       \\
CD-324787 	&   -0.033	     &  0.05	&	0.09   &  0.981       &  3       \\
HD102216  	&   -0.030	     &  0.04	&	0.09   &  1.027       &  3       \\
HD107045  	&   -0.002	     &  0.04	&	0.12   &  0.975       &  3       \\
HD114430  	&   0.003	     &  0.04    &	0.12   &   0.966      &  3       \\
HD116845  	&   -0.007	     &  0.04    &	0.1    &  0.924       &  3       \\
HD122568  	&   -0.018	     &  0.04	&	0.11   &   1.253      &  3       \\
HD124461  	&   -0.008	     &  0.04	&	0.14   &   1.097      &  3       \\
HD162157  	&   0.001	     &  0.04	&	0.11   &   0.966      &  3       \\
HD167768  	&   0.001	     &  0.05	&	0.18   &   0.872      &  3       \\
HD168839  	&   -0.022	     &  0.04	&	0.08   &   0.828      &  3       \\
HD171864  	&   -0.015	     &  0.04	&	0.13   &   1.041      &  3       \\
HD176771  	&   0.030	     &  0.04	&	0.16   &   1.003      &  3       \\
HD191716  	&   -0.007	     &  0.04	&	0.12   &   1.02      &  3       \\
HD96627   	&   -0.010	     &  0.04	&	0.16   &   1.259      &  3       \\
HD99783   	&   -0.007	     &  0.04	&	0.12   &   0.959      &  3       \\
HD116948  	&   0.013	     &  0.04	&	0.11   &    -     &  3       \\
HD162984  	&   -0.035	     &  0.04	&	0.11   &   0.945      &  3       \\

 \hline 
\label{tab:C2019}
\end{tabular}
\end{table}

\begin{table}
\setlength{\tabcolsep}{2pt}
\renewcommand{\arraystretch}{1.}
\centering
\caption{ 
Binary classification of low-mass red giant stars reported in this paper with a single CORALIE or FEROS epoch spectra, including the name of the observed star, the astrometric excess noise (AEN) measurements from $Gaia$ eDR3, the difference between the CORALIE/FEROS RV and the DR2 RV measurements ($\Delta$RV), the RV error from $Gaia$ DR2 (eRV) and the classification number.
}
\centering
\begin{tabular}{lcccccc}
\hline\hline
\vspace{-0.3cm} \\
Star         &     $\Delta$RV    & eRV            &  AEN    &  RUWE         & Category  \\
             &   (km\,s$^{-1}$)  &  (km\,s$^{-1}$)  &(mas)   &           &  \\
\hline \vspace{-0.3cm} \\

HD2259          &   1.36   &     0.50 &   0.45    &      2.695   &  1  \\
HD2132          &   6.27   &     1.47 &   1.90    &      11.862   &  1 (Binary)  \\
HD6254	        &   -1.61	&     0.58 &  0.50    &       2.6   &  1  \\
HD74686	        &   10.23	&     2.45 &  3.42    &       19.337   &  1  \\
HD83674	        &   -2.99	&     0.27 &  0.48    &       2.651   &  1  \\
HD8410	        &   9.41	&     0.53 &  1.33    &       9.195   &  1 (Binary) \\
HD9525	        &   -5.40	&     1.04 &  3.00    &       6.283   &  1 (Binary)  \\
HD10537	        &   -8.28	&     0.15 &  3.61    &       5.214   &  1  \\
HD220864        &   1.82	&     0.15 &  0.63    &       3.578   &  1  \\
HD29821	        &   -8.29	&     0.27 &  0.53    &       3.027   &  1  \\
HD21340	        &   -0.53	&     0.18 &  0.50    &       2.95   &  1  \\
HD169767 &  -11.6       &  0.7      & 0.907  & 2.618 & 1 \\
HD4145	        &   0.90	&     0.15 &  0.20    &       0.993   &  2  \\
HD5676	        &   4.18	&     0.43 &  0.32    &       2.104   &  2  \\
HD6030	        &   1.26	&     1.55 &  0.29    &       1.984   &  2  \\
HD22792	        &   -2.63	&     0.17 &  0.17    &       1.229   &  2  \\
HD18278	        &   -0.29	&     0.23 &  1.04    &       5.464   &  2  \\
HD36787	        &   -0.24	&     1.18 &  0.70    &       3.83   &  2 (Binary)  \\
HD6019	        &   -0.53	&     0.27 &  0.65    &       3.323   &  2  \\
TYC5464-59-1    &   -0.22	&     2.09 &  0.89    &       5.437   &  2  \\
HD10268	        &   -0.22	&     0.15 &  0.28    &       1.719   &  2  \\
HD193937	    &   -0.18	&     0.18 &  0.16    &       1.354   &  2  \\
HD93410  &  0.18   & 0.17   & 0.133  & 1.06  & 2 \\
HD219263	    &    0.11	&     0.18 &  0.32    &       1.212   &  3  \\
HD745	        &   0.08	&     0.18 &  0.17    &       1.05   &  3  \\
HD3303	        &   -0.04	&     0.17 &  0.15    &       0.888   &  3  \\
HD16665	        &   -0.36	&     0.17 &  0.15    &       1.154   &  3  \\
HD27956	        &   -0.24	&     0.12 &  0.16    &       1.089   &  3  \\
HD192232	    &   -0.06	&     0.13 &  0.15    &       0.986   &  3  \\
HD213986	    &   -0.13	&     0.16 &  0.18    &      0.914    &  3  \\
HD220096 	&   -0.35 	     &  0.17 &	0.31 &  0.926   &  3       \\
HD179120 	&   -0.37 	     &  0.13 &	0.11 &  0.995   &  3       \\
HD187562 	&   -0.87 	     &  0.25 &	0.07 &  0.982   &  3       \\
HD188476 	&   -0.35 	     &  0.14 &	0.11 &  0.987   &  3       \\
HD189365 	&   -0.12 	     &  0.14 &	0.11 &  0.96   &  3       \\
HD193407 	&   -0.08 	     &  0.15 &	0.12 &  1.013   &  3       \\
HD195189 	&   -0.09 	     &  0.15 &	0.13 &  0.966   &  3       \\
HD197790 	&   -0.21 	     &  0.14 &	0.13 &  1.033   &  3       \\
HD202704 	&   -0.17 	     &  0.18 &	0.14 &  1.027   &  3       \\
HD203086 	&   -0.23 	     &  0.15 &	0.13 &  1.044   &  3       \\
HD205588 	&   -0.18 	     &  0.17 &	0.12 &  1.005   &  3       \\
HD207920 	&   -0.38 	     &  0.16 &	0.12 &  0.937   &  3       \\
HD209154 	&   -0.33 	     &  0.15 &	0.13 &  0.997   &  3       \\
HD219026 	&   -0.30 	     &  0.14 &	0.16 &  1.149   &  3       \\
HD223700 	&   -0.20 	     &  0.14 &	0.13 &  1.13   &  3       \\
HD4737   	&   -0.30 	     &  0.15 &	0.16 &  1.093   &  3       \\
HD102888 & -0.08   & 0.15   & 0.107  & 0.889 & 3 \\
HD221214 &  0.03   & 0.16   & 0.118  & 0.957 & 3 \\
HD84257  & -0.03   & 0.16   & 0.121  & 0.972 & 3 \\
HD90317  &  0.01  & 0.18   & 0.128  & 0.991 & 3 \\

 \hline 
\label{tab:C20191}
\end{tabular}
\end{table}

\section{Conclusion and future prospects}
\label{conclusion}

Based on the \gaia DR2 color-magnitude diagram, we select a sample of 211 low-mass RG candidates within 200\,pc in order to search for low-mass RGB + MS binary systems with orbital periods between 100 and 900 days, which are potential progenitors of wide binary sdB systems.
From the literature, we found 24 such systems. Furthermore, nine long-period (248 to 898 days) binary systems were identified with orbital parameters from DR3.
In addition, we obtained CORALIE spectroscopic data for 82 stars and single-epoch FEROS archive spectra for 6 stars among this sample. We combined the resulting CORALIE and FEROS RVs with the DR2 and DR3 radial velocity measurements, as well as AEN and RUWE values, to search for potential binary candidates. From these combined datasets we defined three different categories based on their likelihood of being a binary candidate. 
The first group includes binary candidates since they have a large RV variation and a high AEN.
The second group contains likely binary candidates since they show either high RV or have AEN values above the threshold. 
The third group includes the stars that do show neither RV nor AEN variations, which accounts for 53\% of the observed sample, or 47 stars. 
The observed sample demonstrates that 47\% of the observed stars (41 out of 88) most likely have binary signatures.
This result is not unexpected considering that about half of low-mass main sequence stars are found in binary systems \citep{Raghavan2010}.

In future work, we will continue observing the selected binary candidates to confirm their binarity and constrain their orbital parameters. 
Furthermore, the spectra that we obtained will be used to derive accurate atmospheric parameters, including the effective temperature, surface gravity, rotational velocity, microturbulence and metallicity. Also, we will derive accurate radii and masses by deriving the radius from an SED fit for the stars that are presented in this work.
We note that our sample might suffer from possible contamination from the red clump (RC) stars, and this contamination can be up to 30\% as shown by \citet{girardi2016}. Removing this contamination to define the low-mass RG stars is not easy since they cannot be distinguished by either color-magnitude diagrams or spectroscopic surveys \citep[e.g.][]{MH2017}. The way to distinguish these two groups is to use asteroseismology. For instance, \citet{bedding2011} showed that the seismic parameter called period spacings can be used to distinguish RGB stars, burning only hydrogen in the shell, from RC stars that also burn helium in their cores. In the future, we will make use of this advanced technique using high-precision photometry for the targets that have been observed during TESS and K2 missions. 
There are several ongoing large-scale, ground-based surveys focused on obtaining high-resolution spectroscopy, including GALAH (12 $\leq$ V $\leq$ 14), SDSS V (11  $\leq$ H  $\leq$ 14) and $Gaia$-ESO (variable brightness ranges). These surveys will be important to constrain the orbital parameters of promising targets that are classified as 1 and 2 in this work.

\begin{acknowledgements}

M.U. acknowledges financial support from CONICYT Doctorado Nacional in the form of grant number No: 21190886 and ESO studentship program. M.V. acknowledges support from the grant FONDECYT REGULAR No: 1211941.
J.V. acknowledges support from the Grant Agency of the Czech Republic (GA\v{C}R 22-34467S). The Astronomical Institute Ond\v rejov is supported by the project RVO:67985815. A.B. acknowledges support for this project from the
European Union’s Horizon 2020 research and innovation
program under grant agreement No. 865932-ERC-SNeX. Based on observations collected with the CORALIE echelle spectrograph on the 1.2-m Euler Swiss telescope at La Silla Observatory under the program allocated by the Chilean Time Allocation Committee (CNTAC), no: CN2019A-58, CN2019B-87, CN2020B-77, CN2022A-82.
This work has made use of data from the European Space Agency (ESA) mission $Gaia$ (https://www.cosmos.esa.int/$Gaia$), processed by the $Gaia$ Data Processing and Analysis Consortium (DPAC, https://www.cosmos.esa.int/web/$Gaia$/dpac/consortium). Funding for the DPAC
has been provided by national institutions, in particular the institutions participating in the $Gaia$ Multilateral Agreement.

\end{acknowledgements}

\bibliographystyle{aa}
\bibliography{myrefs}

\begin{appendix}
\section{Radial velocity measurements}
\label{Appendix}

\begin{table}
\label{tab:RVs}
\centering
\caption{Radial velocity measurements for all stars that are presented in this work.}
\centering
\begin{tabular}{lccccc}
\hline\hline
\vspace{-0.3cm} \\
Star & JD         &  RV                  & RV Error        & Instrument  \\
     & -2450000   &    (km\,s$^{-1}$)    & (km\,s$^{-1}$)  &        \\
\hline \vspace{-0.3cm} \\

    
HD201013&8650.86 & -20.16  & 0.01  & CORALIE \\
        &9538.56 & -25.73  & 0.01  & CORALIE \\
        
HD204381&8650.82 & -20.96  & 0.01  & CORALIE \\
        &9538.57 & -20.99  & 0.01  & CORALIE \\

HD206005&8650.85 & -9.75  & 0.01  & CORALIE \\
        &9539.52 & -9.26  & 0.01  & CORALIE \\

HD214941&8650.91 & 5.07  & 0.01  & CORALIE \\
        &9538.57 & 1.63  & 0.01  & CORALIE \\

HD217614&8650.79 & -13.93  & 0.01  & CORALIE \\
        &9538.56 & -24.69  & 0.01  & CORALIE \\

CD-324787    &  8649.47     &   0.7677     &   0.0053     &  CORALIE \\   
&9673.49	 &    0.7348	  &   0.0048    &  CORALIE \\
                                                                          
CD-397574    &  8650.60     &   6.0176     &   0.0043     &  CORALIE \\   
&9673.67	 &    5.9716	  &   0.0039    &  CORALIE \\                 

HD102216     &  8649.58     &   17.9042     &   0.004     &  CORALIE \\   
&9673.70	 &    17.8746	  &   0.0031    &  CORALIE \\                 

HD102805     &  8651.54     &   -11.5754     &   0.0039     &  CORALIE \\ 
&9673.64	 &    -11.0578	  &   0.0032    &  CORALIE \\                 

HD107045     &  8650.57     &   -11.6511     &   0.0033     &  CORALIE \\ 
&9673.68	 &    -11.6529	  &   0.0032    &  CORALIE \\                 

HD112521     &  8651.57     &   7.0568     &   0.0042     &  CORALIE \\   
&9673.73	 &    10.7717	  &   0.0033    &  CORALIE \\                 

HD114430     &  8651.58     &   -15.8567     &   0.0045     &  CORALIE \\ 
&9673.72	 &    -15.8535	  &   0.0035    &  CORALIE \\                 

HD116338     &  8651.59     &   -30.5275     &   0.0043     &  CORALIE \\ 
&9673.73	 &    -29.1692	  &   0.0031    &  CORALIE \\                 

HD116845     &  8650.63     &   -6.3724     &   0.0036     &  CORALIE \\  
&9673.74	 &    -6.3798	  &   0.0038    &  CORALIE \\                 

HD116948     &  8651.56     &   -18.5105     &   0.0039     &  CORALIE \\ 
&9673.75	 &    -18.4976	  &   0.0032    &  CORALIE \\                 

HD119483     &  8649.55     &   11.7157     &   0.011     &  CORALIE \\   
&9673.76	 &    11.7758	  &   0.003     &  CORALIE \\                 

HD120144     &  8650.65     &   -0.6528     &   0.0031     &  CORALIE \\  
&9673.76	 &    -0.6394	  &   0.0032    &  CORALIE \\                 

HD122568     &  8651.61     &   17.9469     &   0.0042     &  CORALIE \\  
&9673.78	 &    17.9291	  &   0.0031    &  CORALIE \\                 

HD124461     &  8650.67     &   -40.6677     &   0.0029     &  CORALIE \\ 
&9673.79	 &    -40.6757	  &   0.0032    &  CORALIE \\                 

HD136350     &  8649.64     &   3.7816     &   0.0048     &  CORALIE \\   
&9673.81	 &    -6.1007	  &   0.0034    &  CORALIE \\                 

HD139137     &  8650.69     &   -39.984     &   0.0061     &  CORALIE \\  
&9673.82	 &    -39.0291	  &   0.0068    &  CORALIE \\                 

HD149649     &  8651.64     &   7.936     &   0.005     &  CORALIE \\     
&9673.82	 &    7.9822	  &   0.0035    &  CORALIE \\

HD157527     &  8649.60     &   -52.1753     &   0.0044     &  CORALIE \\ 
&9673.85	 &    -52.4809	  &   0.0034    &  CORALIE \\

HD162157     &  8651.68     &   -21.0394     &   0.0042     &  CORALIE \\ 
&9673.84	 &    -21.0381	  &   0.0032    &  CORALIE \\

HD162984     &  8649.62     &   -9.9796     &   0.0046     &  CORALIE \\  
&9673.85	 &    -10.0147	  &   0.0032    &  CORALIE \\

HD167768     &  8650.81     &   1.6223     &   0.0053     &  CORALIE \\   
&9673.89	 &    1.6237	  &   0.0048    &  CORALIE \\

\hline 
\end{tabular}
\end{table}

\begin{table}
\label{tab:RVs2}
\centering
\caption{Radial velocity measurements for all stars that are presented in this work.}
\centering
\begin{tabular}{lccccc}
\hline\hline
\vspace{-0.3cm} \\
Star & JD         &  RV                  & RV Error        & Instrument  \\
     & -2450000   &    (km\,s$^{-1}$)    & (km\,s$^{-1}$)  &        \\
\hline \vspace{-0.3cm} \\
HD167936     &  8651.70     &   -21.1005     &   0.0065     &  CORALIE \\ 
&9673.86	 &    -37.3306	  &   0.0057    &  CORALIE \\

HD168839     &  8650.89     &   14.4398     &   0.0032     &  CORALIE \\  
&9673.87	 &    14.4181	  &   0.0028    &  CORALIE \\

HD170105     &  8651.66     &   30.6076     &   0.0038     &  CORALIE \\  
&9673.86	 &    31.6746	  &   0.0031    &  CORALIE \\

HD171864     &  8650.70     &   -17.2808     &   0.0027     &  CORALIE \\ 
&9673.87	 &    -17.2956	  &   0.0029    &  CORALIE \\ 

HD176771     &  8649.70     &   -10.2239     &   0.006     &  CORALIE \\  
&9673.88	 &    -10.1944	  &   0.0036    &  CORALIE \\

HD191716     &  8650.80     &   -7.8355     &   0.0029     &  CORALIE \\  
&9673.88	 &    -7.8428	  &   0.0031    &  CORALIE \\ 

HD305357     &  8650.52     &   -0.7367     &   0.0057     &  CORALIE \\  
&9673.58	 &    3.5397	  &   0.0045    &  CORALIE \\

HD71464      &  8649.50     &   16.7708     &   0.0088     &  CORALIE \\  
&9673.53	 &    39.4804	  &   0.0049    &  CORALIE \\ 

HD85885      &  8650.55     &   27.0625     &   0.0063     &  CORALIE \\  
&9673.56	 &    26.8646	  &   0.0052    &  CORALIE \\

HD96627      &  8651.47     &   3.2028     &   0.0036     &  CORALIE \\   
&9673.60	 &    3.1933	  &   0.0032    &  CORALIE \\ 

HD99783      &  8649.60     &   12.5647     &   0.005     &  CORALIE \\   
&9673.61	 &    12.5581	  &   0.0034    &  CORALIE \\

HD99891      &  8651.50     &   -14.3367     &   0.004     &  CORALIE \\  
&9673.62	 &    -14.4727	  &   0.0032    &  CORALIE \\                 

\hline 
\end{tabular}
\end{table}

\end{appendix}

\end{document}